# A protocol for information-driven antibody-antigen modelling with the HADDOCK2.4 webserver.

Francesco Ambrosetti[a], Zuzana Jandova[a] and Alexandre M.J.J. Bonvin[a,*]

[a]Computational Structural Biology Group, Bijvoet Centre for Biomolecular Research, Faculty of Science - Chemistry, Utrecht University, Padualaan 8, 3584 CH Utrecht, the Netherlands.

*Phone: +31.30.2533859, Email: a.m.j.j.bonvin@uu.nl

### i. Summary

In the recent years, therapeutic use of antibodies has seen a huge growth, due to their inherent proprieties and technological advances in the methods used to study and characterize them. Effective design and engineering of antibodies for therapeutic purposes are heavily dependent on knowledge of the structural principles that regulate antibody-antigen interactions. Several experimental techniques such as X-ray crystallography, cryo-electron microscopy, NMR or mutagenesis analysis can be applied, but these are usually expensive and time consuming. Therefore computational approaches like molecular docking may offer a valuable alternative for the characterisation of antibody-antigen complexes.

Here we describe a protocol for the prediction of the 3D structure of antibody-antigen complexes using the integrative modelling platform HADDOCK. The protocol consists of: 1) The identification of the antibody residues belonging to the hyper variable loops which are known to be crucial for the binding and can be used to guide the docking; 2) The detailed steps to perform docking with the HADDOCK 2.4 webserver following different strategies depending on the availability of information about epitope residues.

# 1. Introduction

Antibodies are key players of the immune response against external treats. They consist of two couples of identical polypeptide chains namely light (L) and heavy (H) chain kept together by disulphide bonds [1]. For both heavy and light chains it is possible to identify a variable domain, one constant domain for the light chain, or three or more for the heavy chain. The variable domains of the light (VL) and heavy chain (VH) are involved in the antigen recognition [2]. More specifically, the VL and VH regions contain the so-called Complementarity Determining Regions (CDRs), 3 for the VL and 3 for the VH, which display the highest level of variability and, in most of the cases, directly participate to the interaction with the antigen [3, 4]. Each CDR harbours one loop named hyper variable loop (HV) (six loops in total) which are crucial for the recognition of the cognate antigen and therefore offer a reasonable proxy of the antibody binding interface (paratope).

Over the recent years, the development and application of antibodies for therapeutic purposes has been experiencing a significative boost [5] due to their high specificity and affinity towards specific targets (antigens) and to the advancements in the methodologies applied to produce and characterize them. Knowledge of the relationship between antibody and antigen residues and the way they interact is of paramount importance in order to elaborate effective strategies for their design and engineering [6, 7]. Structural information about their specific interactions can be obtained by using different experimental techniques such as X-ray crystallography, cryo-electron microscopy and Nuclear Magnetic Resonance (NMR). These can provide high resolution information but can also be time consuming, expensive, requiring high amounts of purified samples, and as such not always applicable. Valuable alternatives to the traditional experimental approaches are offered by computational methods which are faster and cheaper. One of them is molecular docking that aims at predicting the structure of a complex starting from the free structures of its components [8]. The process involves two stages: The first one (*sampling*) consists of generating (tens of) thousands of different possible conformations, which are subsequently scored and ranked during the second stage *(scoring)* in order to identify the models that have a higher probability of representing the native structure.

Both sampling and scoring can benefit from the use of available information about e.g. binding interfaces, contacts or other types of information, to either bias the generation of models and/or select the models which are consistent with the provided information [9].

We have recently published an exhaustive comparison of the ability of four well established docking software namely, ClusPro [10], HADDOCK [11], LightDock [12] and ZDOCK [13], in predicting antibody-antigen complexes [14]. All of them can make use of information about the binding interface following different strategies. In this chapter we illustrate how the information about the antibody HV loops and the antigen binding interface (epitope) can be used in HADDOCK, which we demonstrated to be the most accurate for this specific task [14]. The described protocol includes the following steps: Preparation of the structures, selection of residues to use in order to drive the docking, setup of the docking run using the HADDOCK 2.4 web portal and finally analysis of the results.

## 2. Overview

This section describes the different steps to follow to perform the docking of an antibody-antigen complex with HADDOCK.

HADDOCK is an information-driven docking algorithm which can make use of different types of information about the binding interface of two molecules in order to drive the docking [15, 16]. This information can be derived from various experimental sources such as NMR, hydrogen/deuterium exchange, mutagenesis analysis [17, 18] but it can also be extracted from bioinformatic predictions about putative binding sites [19] or co-evolving residues [20]. The information is encoded in the form of distance restraints (ambiguous or unambiguous) which are used both for sampling and scoring to guide the docking process.

In the case of antibodies, as aforementioned, the residues belonging to the HV loops usually are directly involved in the interaction with the antigen and therefore represent a valuable information in order to investigate the antibody-antigen interaction. Nevertheless, on average an antibody uses only around 20-33% of the CDRs residues to engage the antigen [3] therefore depending on the task more precise information can be required. For this reason several computational paratope prediction methods have been developed over the years [21–24]. As for the antigen side, the identification of the epitope is far more challenging and despite the efforts of the community [25–32] none of the current epitope prediction methods show a reasonable accuracy. Their application for molecular docking is therefore largely limited. However, often some experimental evidence pointing to the epitope region on the antigen is available, which can be used to drive the docking with HADDOCK.

## 2.1. HADDOCK docking protocol

The docking in HADDOCK follows three different stages namely:

- Rigid-body energy minimization (it0)
- Semi-flexible refinement by simulated annealing in torsional angle space (it1)
- Final refinement in Cartesian space (itw), with or without explicit solvent.

In the first stages (it0), structures are considered as rigid, separated in spaces and randomly rotated to avoid any bias derived from the starting orientation. Then a rigid-body minimization, driven by the provided data usually transformed into ambiguous distance restraints, is performed allowing the structures to rotate and translate. Despite being rigid, this stage can accommodate ensembles of structures, effectively allowing the docking from various conformations. The second stage (it1) consists of simulated annealing protocol based on short molecular dynamic simulations in torsional angle space (bond lengths and angles are fixed) during which the orientation of side chains first, and then both backbone and side chains of the residues at the interface (defined using a 5Å cut-off) are optimized. Finally, the third stage, performed in Cartesian space, involves either a final energy minimization (default in HADDOCK2.4) or a short molecular dynamics simulation in explicit solvent (default in HADDOCK2.2) in order to further optimize the complex. In all of the three stages available information about the interaction can be included in the form of Ambiguous Interaction Restraints or AIRs (or specific distance restraints, e.g. derived from cross-linking mass spectrometry). These are included as an additional restraining term in the energy function that is minimized, effectively biasing the sampling to account for the information at hand.

After each stage the generated models are scored and ranked according to the stage-specific HADDOCK score (HS) and only the top ranked models (number defined by the user, with a default of 200) move to the next docking step. The HADDOCK score consists of a linear combination of different terms namely: Intermolecular van der Waals ($E_{vdw}$) and electrostatic

($E_{elec}$) energies, an empirical desolvation potential ($E_{desolv}$) [33], the ambiguous interaction restraints energy ($E_{AIR}$) and the buried surface area (BSA). According to the docking stage different weights are associated with each term:

- $HS_{it0}$ = 1.0 $E_{elec}$ + 0.01 $E_{vdw}$ + 1.0 $E_{desolv}$ + 0.1 $E_{AIR}$ - 0.01 BSA
- $HS_{it1}$ = 1.0 $E_{elec}$ + 1.0 $E_{vdw}$ + 1.0 $E_{desolv}$ + 0.1 $E_{AIR}$ - 0.01 BSA
- $HS_{ref}$ = 0.2 $E_{elec}$ + 1.0 $E_{vdw}$ + 1.0 $E_{desolv}$ + 0.1 $E_{AIR}$

## 2.2. Clustering

Once all models have been generated HADDOCK performs a cluster analysis of the final models. By default, this is done based on the fraction of common contacts (FCC) using 0.6 as cutoff [34], but also the interface-ligand root mean square deviation (i-l-RMSD) can be used depending on the user's selection.

## 3. Method

In order to use the HADDOCK2.4 web-server registration is required. This can be done at: https://wenmr.science.uu.nl/auth/register/. To have access to all the functionalities required for this protocol it is necessary to request at least the *Expert* level for HADDOCK (this can be done in your registration page). In the following paragraphs we will illustrate this protocol step by step, using as test case the 4G6M antibody-antigen complex present in the dataset we used in [14]. This complex describes the interaction between the humanized antibody *Gevokizumab* (4G6K) and its cognate antigen, *interleukin-1beta* (4I1B) (both structures and the code used in this protocol are provided in the GitHub repository: https://github.com/haddocking/HADDOCK-antibody-antigen). We will describe two protocols corresponding to the first two scenarios described the our previous work [14], either when no information is available for the antigen binding site, or when some vague information about the epitope is provided to guide the docking.

### 3.1. Installation

In order to follow this protocol, it is necessary to download install all the requested dependencies and ANARCI [35], which is the software we use in order to renumber the antibody and extract the information about the HV loops. The steps described here assume either a Linux or OSX environment. We are using anaconda (https://www.anaconda.com/distribution/) for an easy installation of the dependencies, but it is also possible to install them manually. To run this protocol you will need to install:

1. Python 2.7 (https://www.python.org/downloads/release/python-2713/)
2. HMM 3.3 (http://hmmer.org/) [36]
3. Biopython (https://biopython.org/)
4. Biopandas (http://rasbt.github.io/biopandas/) [37]
5. PDB-tools (https://github.com/haddocking/pdb-tools) [38]
6. ANARCI (http://opig.stats.ox.ac.uk/webapps/newsabdab/sabpred/anarci/) [35]

All the instructions are provided in the README.md file of the GitHub repository.

1. Open a Linux window terminal and download the repository by typing:

```
> git clone https://github.com/haddocking/HADDOCK-antibody-antigen
```

2. Different strategies can be used to install all the dependencies:

    a. (Recommended) If you have anaconda installed following the installation instructions provided in the README.md file type:

    ```
    > cd HADDOCK-antibody-antigen
    > conda env create
    > conda activate Ab-HADDOCK
    > cd anarci-1.3
    > python2.7 setup.py install
    ```

```
> conda deactivate
> cd ..
```

  b. If you do not have anaconda you will need to install all the dependencies separately following the instructions on the README.md file and on the specific websites:

   i. Download and install python 2.7: https://www.python.org/

   ii. Download and install HMMER 3.3: http://hmmer.org/

   iii. Install the required python packages:

```
> cd HADDOCK-antibody-antigen
> pip install -r requirements.txt
```

   iv. Install ANARCI:

```
> cd anarci-1.3
> python2.7 setup.py install
> cd ../..
```

## 3.2. Identification of the Hyper Variable loops

Starting from the antibody structure downloaded from the Protein Data Bank (PDB) [39] we will first identify the residues belonging to the HV loops in order to use them to drive the docking.

The steps are the following:

1. (Optional) Only if you have used anaconda to install dependencies you need to activate the corresponding environment by typing:

```
> conda activate Ab-HADDOCK
```

2. Using a terminal window run ANARCI to renumber the antibody (4G6K) according to the Chothia numbering scheme:

```
> python2.7 ImmunoPDB.py -i 4G6K.pdb -o 4G6K_ch.pdb --scheme c –fvonly –rename --splitscfv
```

The argument *-i*, *-o*, and *--scheme* are related to the input file, output file and to the numbering scheme to be used, respectively. *--fvonly* tells the script to only output the variable domain of the antibody discarding the constant one known to be not involved in the interaction with the antigen. The option *--rename* is used to allow the script to rename the antibody chain IDs in H and L for the heavy and light chain respectively. Finally *--splitscfv* is provided in order to tell the script to split the variable domain of single chain antibodies.

3. Format the antibody structure in order to match the HADDOCK format requirements (see **Note 1**) and extract the new residue number of the HV loop residues:

```
> python ab_haddock_format.py 4G6K_ch.pdb 4G6K-HADDOCK.pdb A > active.txt
```

The first argument is the input PDB file that must be formatted according to the Chothia numbering scheme (see **Note 2**), the second one is the name of the output formatted PDB file and finally the third one is the chain ID to be used in the HADDOCK formatted version of the PDB file. This scripts also outputs a comma separated list of residues associated to the amino acids of the antibody HV loops (see **Figure 1**). In the above command this list is written the *active.txt* file.

4. Add TER and END statements to the HADDOCK-ready PDB file (*4G6K-HADDOCK.pdb*)

```
> pdb_tidy 4G6K-HADDOCK.pdb > oo; mv oo 4G6K-HADDOCK.pdb
```

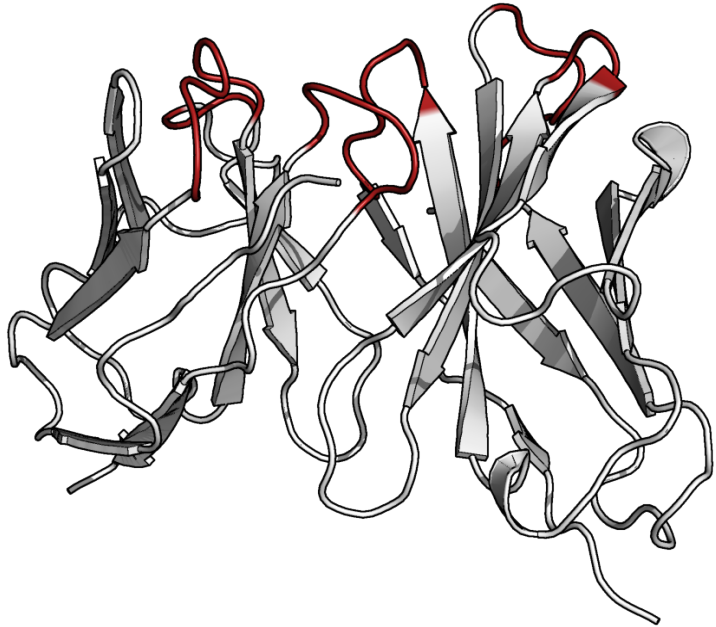

**Figure 1**: Structure of the variable domain of an antibody (PDB code: 4G6K). In red are highlighted the hyper variable loops defined according to the Chothia numbering scheme.

### 3.3. Preparation of the input files

The PDB files used as input in HADDOCK need to be carefully checked in order to fit the HADDOCK format requirements, e.g. removing double occupancies and insertions (relevant in the case of antibodies). For removing double occupancies the script *pdb_selaltloc.py* provided in the PDB-tools repository (https://github.com/haddocking/pdb-tools) [38] can be used, while insertions have been automatically removed using the ab_*haddock_format.py* script as shown in the step 3 of the previous paragraph. Alternatively, to deal with the insertions, it is possible to use the script *pdb_delinsertion.py* of the PDB-tools repository.

### 3.4. Antibody-antigen docking

For the docking we will use the new HADDOCK2.4 web portal: https://wenmr.science.uu.nl/haddock2.4/. To follow this protocol you need to be registered and have at least *Expert level* access.

1. From an Internet browser go to: https://wenmr.science.uu.nl/haddock2.4/ and click on "*Submit a new job*".

2. Provide a job name for the docking run. Here we will use: *4G6M-Ab-Ag*.

3. In the *Molecule 1* section for the entries "*Where is the structure provided?*" and "*Which chain of the structure must be used?*" leave the default options: "*I am submitting it*" and "*All*", respectively. Then click on "*Choose file*" and upload the HADDOCK-formatted PDB file of the antibody created at the stage 3 of paragraph 3.2 (*4G6K-HADDOCK.pdb*). Leave the remaining options to their default values (see **Figure 2**).

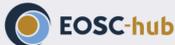

**Figure 2**: Representation of the HADDOCK input page.

4. In the *Molecule 2* section as before leave "*I am submitting it*" and "*All*" for the entries "*Where is the structure provided?*" and "*Which chain of the structure must be used?*", respectively and by clicking on "*Choose file*" upload the antigen PDB file provided in the repository (*4I1B-matched.pdb*). Leave the remaining options to their default values.

5. Click "*Next*" at the bottom of the page (see **Note 3**).

6. In the "*Molecule 1 - parameters*" section copy and paste the list of the HV loop residues from the *active.txt* file created in the step 3 of the previous paragraph and deactivate the option: "*Automatically define passive residues around the active residues*". Leave the other parameters to their default values.

7. In the "*Molecule 2 - parameters*" different strategies need to be followed according to the availability or not of information about the epitope residues:

    a. When no information about the epitope is provided deactivate the option: "*Automatically define passive residues around the active residues*" and enable the option: "*Automatically define surface residues as passive*". In the entry "*If you specified that surface residues will be defined automatically as passive, selection will use the following RSA (relative solvent accessibility) cutoff*" leave 0.40 as cutoff. Leave the other options to their default parameters.

    b. When some information about the epitope is available deactivate the option: "*Automatically define passive residues around the active residues*" and in the field "*Passive residues (surrounding surface residues)*" copy and paste the list of comma separated epitope residues. Just as an example here we will use the residues: *22,46,47,48,64,71,72,73,74,75,82,84,85,86,87,91,92,95,114,116,117* which represent all of the antigen residues within 9Å from the antibody in the complex reference structure (4G6M), filtered by their relative solvent accessibility (≥0.40) upon the removal of the antibody.

8. Click "*Next*" at the bottom of the page.

9. In the "*Sampling parameters*" section, depending on whether you are using information about the epitope or not, it is necessary to tune the sampling parameters:

    a. If you are not using any information about the epitope change increase the sampling by changing "*Number of structures for rigid body docking*", "*Number*

*of structures for semi-flexible refinement*" and "*Number of structures for water refinement*", to 10000, 400 and 400 respectively (see **Note 4**).

  b. If you are providing a loose definition of the epitope leave the entries: "*Number of structures for rigid body docking*", "*Number of structures for semi-flexible refinement*" and "*Number of structures for water refinement*", to 1000, 200 and 200 respectively (the default values).

  Leave the other options unchanged.

10. Click "*Submit*" at the bottom of the page.

11. A page reporting a small summary of your run and the progress bar of the docking job will be loaded. This page refreshes every 30 seconds (see **Note 5**).

12. Depending on the load on the server, the docking process might take between one and several hours. A link with the results will be sent by e-mail. From this page it is possible to have an overview of the docking run, inspect models, download all cluster representatives or download the full archive of the run, including all models from all stages as a *gzipped tar* file (typically a few 100MBs to a few GBs in size, depending on the system size).

13. The result page reports the number of clusters and for the top 10 clusters also the related statistics (HADDOCK score, Size, RMSD, Energies, BSA and Z-score). While the name of the clusters is defined by their size (cluster 1 is the largest, followed by cluster 2 etc..) the top 10 clusters are selected and sorted according to the average HADDOCK score of the best 4 models of each cluster, from the lowest HADDOCK score to the highest. The various models can be directly visualized online by clicking on the eye icon, or downloaded for further analysis.

14. The bottom of the result page, the "*Model Analysis*" section, presents interactive plots. These allow the user to inspect the distributions of HADDOCK score and associated energetic terms as a function of the fraction of common contacts (FCC) and interface RMSD (i-RMSD). FCC and i-RMSD are calculated with respect to the overall top scoring model. The points are color-coded by cluster. Clusters can be turned on and off. Finally, in the "*Cluster analysis*" section the distribution of the energetic terms of the HADDOCK score are shown for each cluster (see **Figure 3**).

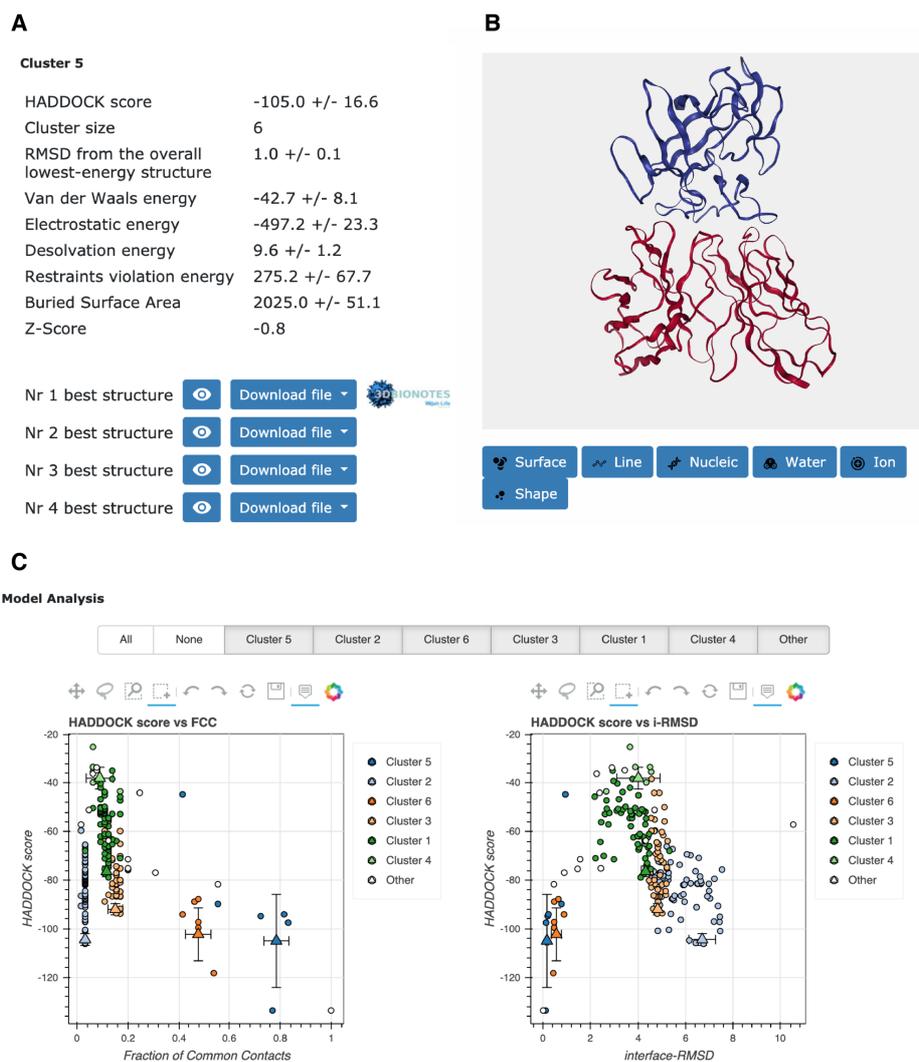

**Figure 3**: Illustration of the result page of HADDOCK2.4. **A)** The result page reports different statistics for the top 10 HADDOCK clusters. Those consist of the HADDOCK score, various energetic and geometric terms and the positional interface ligand RMSD (i-l-RMSD) of the top 4 clustered models with respect to the model with the overall lowest HADDOCK score. Their average and standard deviation are calculated from the top 4 ranked models of each cluster. The z-score expresses the number of standard deviations that a cluster score is from the average of all clusters scores. **B)** The result page allows to directly visualize the models in an interactive window. **C)** View of two of the interactive plots provided in the results page: HADDOCK score versus fraction of common contacts (FCC) (left) and interface RMSD (right) with respect to the top ranked model.

In this particular example, it is possible to compare the docking results with the reference structure (4G6M) for example by downloading all the cluster representative models. Those models can be directly downloaded from the result page or can be found in the directory containing all the results. In the CAPRI (Critical PRediction of Interactions) [40] experiment, one of the parameters used is the Ligand RMSD (l-RMSD) which is calculated by superimposing the structures onto the backbone atoms of the receptor (the antibody in this case) and calculating the RMSD on the backbone residues of the ligand (the antigen) (see **Figure 4**). To calculate the l-RMSD it is possible to either use the software Profit (http://www.bioinf.org.uk/software/profit/) or Pymol (https://pymol.org/2/). It is simpler if the reference PDB structure has the same sequence numbering and chain IDs as the HADDOCK-models. As an example, in the GiHub repository we provide the formatted reference PDB (4G6M-matched). Starting from this structure it is possible to calculate the l-RMSD with Pymol by running the following code in the Pymol console:

```
> alter all, segi=''
> align cluster5_1 and chain A, 4G6M-matched and chain A, cycles=0
> rms_cur cluster5_1 and chain B, 4G6M-matched
```

As example here we used the best structure of the best cluster but the action can be repeated for all the HADDOCK-models (see **Note 6**).

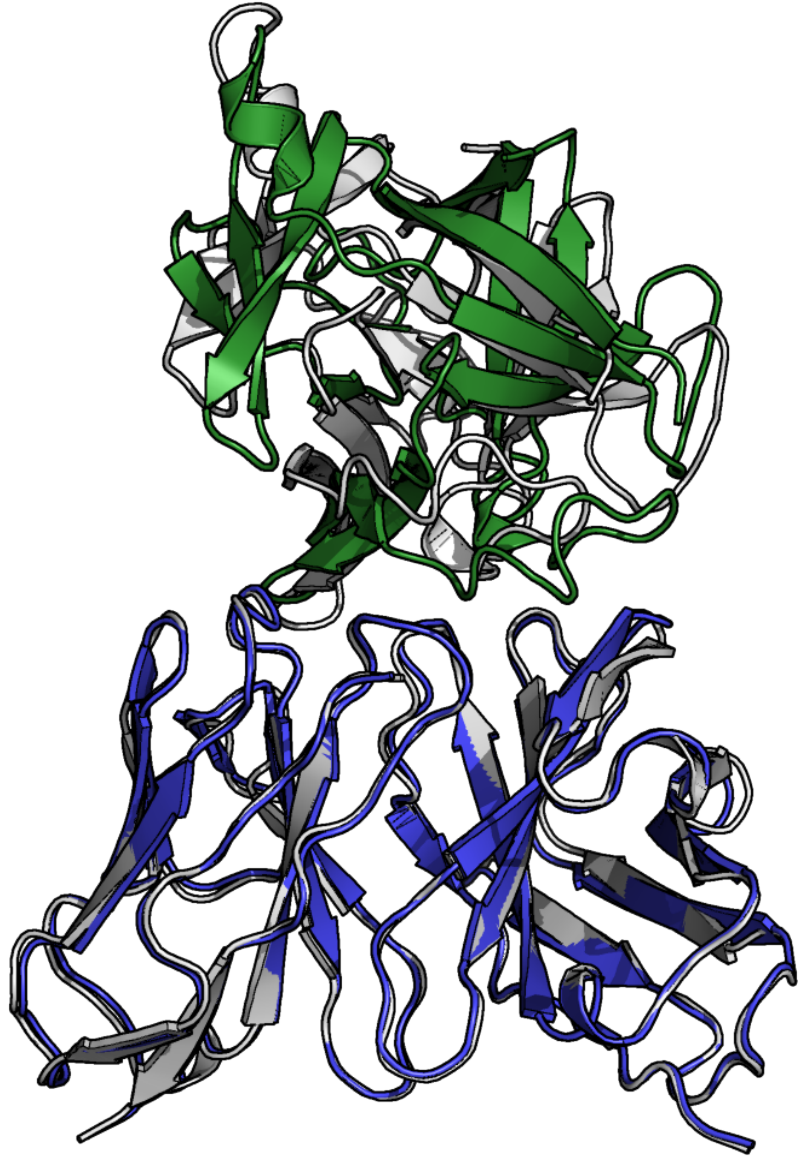

**Figure 4**: Comparison of the top ranked model of the top ranked HADDOCK cluster with the reference structure (4G6M). The structures are superimposed on the antibody (ligand RMSD = 8.0Å and interface RMSD = 2.7Å) (see **Note 6**). The antibody and the antigen of the docking model are represented in blue and green, respectively. The reference structure is reported in grey. The docking was driven by using the hypervariable loops (HV) of the antibody and the epitope residues as defined in the text (paragraph 3.4 step 7b).

## Acknowledgments

This work is supported by the European Union Horizon 2020 BioExcel (grant# 675728 and 823830) and EOSC-hub (grant# 777536) projects.

**Notes**

1. The PDB file of the antibody has to be formatted to match the HADDOCK format requirements. Here the antibody is treated as a single chain molecule with no insertions and no overlapping numbering. To validate the PDB format it is also possible to use the *pdb_validate.py* of the PDB-tools directory (https://github.com/haddocking/pdb-tools).

2. The PDB file used to extract the hyper variable loops must be numbered according to the Chothia numbering scheme (different schemes would lead to the wrong identification of the HV residues) and the heavy and light chains must have appropriate IDs: H and L, respectively.

3. After the submission of the two molecule they are validated by running through Molprobity [41]. This also allows for the automatic definition of the histidine protonation state.

4. In general, the less information is provided the larger the sampling should be. In this specific case the increase of the sampling is necessary in order to enable HADDOCK to sample the entire surface of the antigen.

5. The user will be notified by email upon successful submission and after completion of the run. It is therefore not required to leave the web page open all the time.

6. Given the intrinsic chaotic nature of the computations performed by HADDOCK, which distributes computations on a worldwide grid of computers, differences in scores and accordingly cluster rankings might be expected unless the runs are performed on exactly the same hardware, operating system and using the same executable. The overall results in term of models should however be consistent.

# References

1. Narciso JET, Uy IDC, Cabang AB, et al (2011) Analysis of the antibody structure based on high-resolution crystallographic studies. N Biotechnol 28:435–447. https://doi.org/10.1016/j.nbt.2011.03.012

2. Novotný J, Bruccoleri R, Newell J, et al (1983) Molecular anatomy of the antibody binding site. J Biol Chem 258:14433–14437

3. Sela-Culang I, Kunik V, Ofran Y (2013) The Structural Basis of Antibody-Antigen Recognition. Front Immunol 4:302. https://doi.org/10.3389/fimmu.2013.00302

4. MacCallum RM, Martin ACR, Thornton JM (1996) Antibody-antigen interactions: Contact analysis and binding site topography. J Mol Biol 262:732–745. https://doi.org/10.1006/jmbi.1996.0548

5. Kaplon H, Reichert JM (2019) Antibodies to watch in 2019. MAbs 11:219–238. https://doi.org/10.1080/19420862.2018.1556465

6. Morea V, Lesk AM, Tramontano A (2000) Antibody modeling: Implications for engineering and design. Methods 20:267–279. https://doi.org/10.1006/meth.1999.0921

7. Norman RA, Ambrosetti F, Bonvin AMJJ, et al (2019) Computational approaches to therapeutic antibody design: established methods and emerging trends. Brief Bioinform. https://doi.org/10.1093/bib/bbz095

8. Moreira IS, Fernandes PA, Ramos MJ (2010) Protein-protein docking dealing with the unknown. J Comput Chem 31:317–342. https://doi.org/10.1002/jcc.21276

9. Rodrigues JPGLM, Bonvin AMJJ (2014) Integrative computational modeling of protein interactions. FEBS J. 281:1988–2003

10. Kozakov D, Hall DR, Xia B, et al (2017) The ClusPro web server for protein-protein docking. Nat Protoc 12:255–278. https://doi.org/10.1038/nprot.2016.169

11. Dominguez C, Boelens R, Bonvin AMJJ (2003) HADDOCK: A Protein−Protein Docking Approach Based on Biochemical or Biophysical Information. J Am Chem Soc 125:1731–1737. https://doi.org/10.1021/ja026939x

12. Jiménez-García B, Roel-Touris J, Romero-Durana M, et al (2018) LightDock: A new multi-scale approach to protein-protein docking. Bioinformatics 34:49–55. https://doi.org/10.1093/bioinformatics/btx555

13. Chen R, Weng Z (2002) Docking unbound proteins using shape complementarity, desolvation, and electrostatics. Proteins Struct Funct Genet 47:281–294. https://doi.org/10.1002/prot.10092

14. Ambrosetti F, Jiménez-García B, Roel-Touris J, Bonvin AMJJ (2020) Modeling Antibody-Antigen Complexes by Information-Driven Docking. Structure 28:119-129.e2. https://doi.org/10.1016/j.str.2019.10.011

15. Melquiond ASJ, Bonvin AMJJ (2010) Data-driven docking: Using external information to spark the biomolecular rendez-vous. In: Protein-Protein Complexes: Analysis, Modeling and Drug Design

16. Karaca E, Bonvin AMJJ (2013) Advances in integrative modeling of biomolecular complexes. Methods

17. Lim XX, Chandramohan A, Lim XYE, et al (2017) Epitope and Paratope Mapping

Reveals Temperature-Dependent Alterations in the Dengue-Antibody Interface. Structure 25:1391-1402.e3. https://doi.org/10.1016/j.str.2017.07.007

18. Fontayne A, De Maeyer B, De Maeyer M, et al (2007) Paratope and epitope mapping of the antithrombotic antibody 6B4 in complex with platelet glycoprotein Ibα. J Biol Chem 282:23517–23524. https://doi.org/10.1074/jbc.M701826200

19. de Vries SJ, Bonvin AMJJ (2011) CPORT: A Consensus Interface Predictor and Its Performance in Prediction-Driven Docking with HADDOCK. PLoS One 6:e17695. https://doi.org/10.1371/journal.pone.0017695

20. Hopf TA, Schärfe CPI, Rodrigues JPGLM, et al (2014) Sequence co-evolution gives 3D contacts and structures of protein complexes. Elife 3:. https://doi.org/10.7554/eLife.03430

21. Ambrosetti F, Olsen TH, Olimpieri PP, et al (2020) proABC-2: PRediction Of AntiBody Contacts v2 and its application to information-driven docking. bioRxiv 2020.03.18.967828. https://doi.org/10.1101/2020.03.18.967828

22. Liberis E, Velickovic P, Sormanni P, et al (2018) Parapred: Antibody paratope prediction using convolutional and recurrent neural networks. Bioinformatics 34:2944–2950. https://doi.org/10.1093/bioinformatics/bty305

23. Krawczyk K, Baker T, Shi J, Deane CM (2013) Antibody i-Patch prediction of the antibody binding site improves rigid local antibody-antigen docking. Protein Eng Des Sel 26:621–629. https://doi.org/10.1093/protein/gzt043

24. Kunik V, Ashkenazi S, Ofran Y (2012) Paratome: an online tool for systematic identification of antigen-binding regions in antibodies based on sequence or structure. Nucleic Acids Res 40:W521-4. https://doi.org/10.1093/nar/gks480

25. Sela-Culang I, Ashkenazi S, Peters B, Ofran Y (2015) PEASE: Predicting B-cell epitopes utilizing antibody sequence. Bioinformatics 31:1313–1315. https://doi.org/10.1093/bioinformatics/btu790

26. Krawczyk K, Liu X, Baker T, et al (2014) Improving B-cell epitope prediction and its application to global antibody-antigen docking. Bioinformatics 30:2288–2294. https://doi.org/10.1093/bioinformatics/btu190

27. Jespersen MC, Peters B, Nielsen M, Marcatili P (2017) BepiPred-2.0: Improving sequence-based B-cell epitope prediction using conformational epitopes. Nucleic Acids Res 45:W24–W29. https://doi.org/10.1093/nar/gkx346

28. Qi T, Qiu T, Zhang Q, et al (2014) SEPPA 2.0—more refined server to predict spatial epitope considering species of immune host and subcellular localization of protein antigen. Nucleic Acids Res 42:W59–W63. https://doi.org/10.1093/nar/gku395

29. Liang S, Zheng D, Standley DM, et al (2010) EPSVR and EPMeta: Prediction of antigenic epitopes using support vector regression and multiple server results. BMC Bioinformatics 11:381. https://doi.org/10.1186/1471-2105-11-381

30. Kringelum JV, Lundegaard C, Lund O, Nielsen M (2012) Reliable B Cell Epitope Predictions: Impacts of Method Development and Improved Benchmarking. PLoS Comput Biol 8:e1002829. https://doi.org/10.1371/journal.pcbi.1002829

31. Rubinstein ND, Mayrose I, Martz E, Pupko T (2009) Epitopia: a web-server for predicting B-cell epitopes. BMC Bioinformatics 10:287. https://doi.org/10.1186/1471-2105-10-287

32. Ansari HR, Raghava GP (2010) Identification of conformational B-cell Epitopes in an antigen from its primary sequence. Immunome Res 6:6. https://doi.org/10.1186/1745-7580-6-6

33. Fernández-Recio J, Totrov M, Abagyan R (2004) Identification of protein-protein interaction sites from docking energy landscapes. J Mol Biol. https://doi.org/10.1016/j.jmb.2003.10.069

34. Rodrigues JPGLM, Trellet M, Schmitz C, et al (2012) Clustering biomolecular complexes by residue contacts similarity. Proteins Struct Funct Bioinforma 80:1810–1817. https://doi.org/10.1002/prot.24078

35. Dunbar J, Deane CM (2016) ANARCI: Antigen receptor numbering and receptor classification. Bioinformatics 32:298–300. https://doi.org/10.1093/bioinformatics/btv552

36. Eddy SR (2011) Accelerated profile HMM searches. PLoS Comput Biol. https://doi.org/10.1371/journal.pcbi.1002195

37. Raschka S (2017) BioPandas: Working with molecular structures in pandas DataFrames. J Open Source Softw. https://doi.org/10.21105/joss.00279

38. Rodrigues J, Teixeira JMC, Trellet M, et al (2020) haddocking/pdb-tools: Bug Fix Release. https://doi.org/10.5281/ZENODO.3608327

39. Berman HM, Battistuz T, Bhat TN, et al (2002) The Protein Data Bank. Acta Crystallogr Sect D Biol Crystallogr 58:899–907. https://doi.org/10.1107/S0907444902003451

40. Méndez R, Leplae R, De Maria L, Wodak SJ (2003) Assessment of blind predictions of protein-protein interactions: Current status of docking methods. Proteins Struct Funct Genet 52:51–67. https://doi.org/10.1002/prot.10393

41. Davis IW, Leaver-Fay A, Chen VB, et al (2007) MolProbity: All-atom contacts and structure validation for proteins and nucleic acids. Nucleic Acids Res. https://doi.org/10.1093/nar/gkm216